\documentclass[a4paper,8pt]{article}
\usepackage[cp1251]{inputenc}
\usepackage[russian,english]{babel}
\usepackage{array,longtable}
\usepackage{amssymb, latexsym, amsthm, amsmath, amsxtra}
\usepackage{graphicx}

\graphicspath{{}}
\DeclareGraphicsExtensions{.pdf,.png,.jpg,}

\makeatletter

\begin{document}
\large

\begin{center}
\title{}{\bf On a nonlinear model of the localized vacuum hypothesis, for  solving  the cosmological constant problem.  }
\vskip 1cm

\author{}
 R.K. Salimov \textsuperscript{1}
{}

% author(-s)
\vskip 1cm

{ \textsuperscript{1} Bashkir State University, 450076, Ufa, Russia }

%The name of establishment in which research is executed.
%If authors take an identical place of work, the name of the organization is written once.

\vskip 0.5cm
e-mail: salimovrk@bashedu.ru

\end{center}

\vskip 1cm

{\bf Abstract}
 \par
A new model of oscillators was  suggested, in which an oscillating particle in the minimum energy state has a nonzero velocity. A system consisting of a point material particle and a scalar field described by the nonlinear Klein-Gordon equation has been considered. It has been shown that, when taking into account relativistic effects, in the case of small rest masses of a particle an energy minimum at zero velocity is impossible for such a particle.  It is showed that the behavior of a field in such a system is not stationary and is characterized by the presence of waves emitted and absorbed by the system in the minimum energy state. The system properties having being analyzed, a concept of the localized vacuum was suggested; it was showed that the localized vacuum hypothesis is useful in solving the cosmological constant problem.

 \par
 \vskip 0.5cm


\par
\vskip 1cm

One of the invariant nonlinear differential equations studied most often is the Klein-Gordon equation, the sine-Gordon, in particular. It has a lot of applications in different fields of physics including hydrodynamics, condensed matter physics, field theory, etc. [1-4]. Impurities are normally thought to be stationary when considering equations with different inhomogeneities and impurities. Inhomogeneities and impurities in such problems simulate various defects like the defects in magnetic materials for the the sine-Gordon equation [5-7]. The Klein-Gordon equations are Lorentz-invariant and their solutions have  relativistic effects [8]. For this reason it is quite interesting to study a system of these equations and a point particle described by the relativistic dynamics. A point particle in the given model is a source of  inhomogeneity for the scalar field. Moreover, as is showed in [9], taking relativistic effects in such a system into account results in the emerging of an undamped motion which can be considered as a model of the intrinsic angular momentum or a particle spin [10].

             To consider nonstationary inhomogeneities or defects we assume that inhomogeneity is created by a particle with the mass $m$, the coordinate of which is denoted as $z$, its velocity as $v=\dot z$.  The Hamiltonian of the inhomogeneity interacting with the field u can be written in the form (in the one-dimensional case).
\begin{align}
 H=H_{def}+H_u+H_{int} \label{eq:1}
   \end{align}
   where  $H_{def}$ is the energy of the particle creating the  inhomogeneity.
     \begin{align}
 H_{def}=\frac{m}{\sqrt{(1-v^2)}}  \label{eq:1}
   \end{align}

 $H_u$ is the scalar field energy

    \begin{align}
  H_u= \int\limits_{-\infty}^{\infty} (\frac{u_x^2}{2}+\frac{u_t^2}{2}+V(u)) dx  \label{eq:3}
   \end{align}
$H_{int}$ is the energy of interaction between the scalar field and the particle creating the  inhomogeneity.
 \begin{align}
  H_{int}= \int\limits_{-\infty}^{\infty}  q(z,v,x)W(u) dx  \label{eq:3}
   \end{align}

Function $V(u)$ in expression (3)  was written in the form:
 \begin{align}
  V(u)=\frac{u^2}{2}+\frac{u^4}{2}  \label{eq:3}
   \end{align}

From the Hamiltonian-preserving condition (1),  differentiating it with respect to time, we obtain motion equations for the field and the particle.
Then, integrating
\begin{align}
  \int\limits_{-\infty}^{\infty} u_x u_{xt} dx=-\int\limits_{-\infty}^{\infty} u_{xx} u_{t} dx
\end{align}
And assuming that the motion equation for the field $u$ is satisfied
\begin{align}
 -u_{xx}+u_{tt}+\frac{\partial V}{\partial u}+q\frac{\partial W}{\partial u}=0
   \end{align}
We obtain the motion equation for the particle
\begin{align}
   \int\limits_{-\infty}^{\infty} (W\frac{\partial q}{\partial z}\dot z+W\frac{\partial q}{\partial v}\dot v)dx+\frac{m v\dot v}{(1-v^2)^{(3/2)}}=0
   \end{align}
 Consider next the case of the interaction  $H_{int}$ , where the scope of the potential  $q(z,v,x)$ is limited by the region which can be denoted by the  Heaviside function, and $W=2\cos^2(u/2)$.

\begin{eqnarray}
  q(x,z,v)W(u)= \frac{U_0 2\cos^2(u/2)}{\sqrt{1-v^2}}(\theta(x-(z-l\sqrt{1-v^2})\nonumber\\-\theta(x-(z+l\sqrt{1-v^2})))
  \end{eqnarray}
Here, the relativistic change in the potential at motion of its source, i.e. the particle,  is taken similar to the change in the  electrostatic potential $\phi$  in the relativistic case, e.g. the change in the potential of a moving charge. This means that the value of the potential at motion for an observer at rest becomes larger and its action region becomes narrower. Retardation is neglected in this case. Such an interaction potential results in the equation of motion for the field
\begin{align}
  u_{xx}-u_{tt}=u+2u^3-\nonumber\\\frac{U_0 \sin(u)}{\sqrt{1-v^2}}(\theta(x-(z-l\sqrt{1-v^2})-\theta(x-(z+l\sqrt{1-v^2}))
   \end{align}
 and for the paricle
\begin{align}
 \frac{mv\dot v}{(1-v^2)^{(3/2)}}+\frac{U_0 v \dot v}{(1-v^2)^{3/2}}\int\limits_{z-l\sqrt{1-v^2}}^{z+l\sqrt{1-v^2}}2cos^2(u/2)dx\nonumber\\-\frac{U_0 v\dot v l}{(1-v^2)}2cos^2(u(z-l\sqrt{1-v^2})/2)
  \nonumber\\-\frac{U_0 v \dot v l}{(1-v^2)}2cos^2(u(z+l\sqrt{1-v^2})/2))=\nonumber\\\frac{U_0 \dot z }{\sqrt{1-v^2}}2cos^2(u(z-l\sqrt{1-v^2})/2)
  \nonumber\\-\frac{U_0  \dot z }{\sqrt{1-v^2}}2cos^2(u(z+l\sqrt{1-v^2})/2)
   \end{align}

     The equation has a numerical stationary solution, e.g. for the parameters $U_0=20$, $l=0.5$ [17]. The solution being monotonous, i.e. $u$ decreasing with the growth of  $x-x_0$  and $|u|_{max}<\pi$, is enough for further reasoning. The coordinates of the center of the $z$ potential coincide with the coordinates of the center of the soliton solution $x_0$. Notice also that the value $cos^2(u/2)$ increases monotonously with the growth of $x-x_0$. We further assume that at small rest masses $m$  for a stationary localized particle the energy minimum $H_{def}+H_u+H_{int}$ is achieved. We show then it is not true. To do so we calculate the values of  partial derivatives $$\frac{\partial H}{\partial v} , \frac{\partial^2 H}{\partial v^2}, \frac{\partial H}{\partial z} , \frac{\partial^2 H}{\partial v \partial z},\frac{\partial^2 H}{ \partial z^2}$$
      for $v=0$ and $z=x_0$.  Due to the monotonicity of $cos^2(u/2)$ for the stationary soliton solution for the field $u$, we obtain at rather small $m$ $$\frac{\partial H}{\partial v}=0,\frac{\partial H}{\partial z}=0,\frac{\partial^2 H}{\partial v^2}<0, \frac{\partial^2 H}{\partial v \partial z}=0,\frac{\partial^2 H}{ \partial z^2}>0$$ at $v=0$ , $z=x_0$. That is, for rather small $m$ the Hamiltonian $H_{def}+H_u+H_{int}$ has no local minimum as a function of variables $v$ and $z$ at $v=0$ , $z=x_0$. Paper [17] describes the unstable stationary state in more detail.

      That is, it was showed that the minimum energy of such a system is smaller than the energy of the system of the stationary soliton solution at a stationary inhomogeneity.

Besides, it is important to note that a state with some nonzero localized solution for the field $u$ and a particle being the source of the inhomogeneity and finitely oscillating  about the state, is a minimum energy state. Indeed, if the particle leaves the localized solution region, the energy value $H=H_{def}+H_u+H_{int}$  increases and becomes larger than the energy of the state of the stationary soliton solution at a stationary inhomogeneity. The solution at which the value of the field $u=0$ also has the energy larger than the energy of the state of the stationary soliton solution at a stationary inhomogeneity

Hence, a state with a minimum energy is the state with some nonzero localized solution for the field $u$ and a particle finitely oscillating  about the state. Since the state of the particle is not stationary, the solution for the field $u$ is not stationary as well. There are some oscillations in the neighborhood of the localized solution of the field $u$ but the energy is not emitted away from the localized solution since the energy of the system is minimal. Such oscillations can, in some sense, be considered as a sum of emitted and absorbed waves, or virtual waves, while the particle interacting with the field $u$ is the one interacting with the virtual waves or some local vacuum of such waves. We call the vacuum local since away from the localized solution with a minimum energy the field is $u\to 0$ and the field energy is $V\to 0$. Due to the oscillations the point particle location is, in a sense, “smeared out” in space.  The analogy of complementary oscillations of an electron at the Lamb shift  is relevant for such a system.
It should be noted here that at present quantum electrodynamics is the most precise in its predictions and the most developed quantum field theory. But the cosmological constant problem [11-13] makes the assumption about quantum electrodynamics not being a final theory quite relevant.

A hypothesis of the localized vacuum is suggested to solve the  vacuum catastrophe problem.  To decrease the guess value of the vacuum energy of virtual photons we assume they only exist in a certain region of charged particles. That is, we assume that the density of the energy of zero-point oscillations equals zero for the space without charged particles or far away from them. Physical effects, such as the Lamb shift, explained by the interaction with zero-point oscillations in QED, in the frames of the given hypothesis are explained as a consequence of the local interaction of particles with some scalar field. The energy of zero-point oscillations in an inane space is taken as zero. Then the problem of infinite energy of zero-point oscillations is also solved. The example of a nonlinear relativistic model considered above shows that certain perspectives are possible for the hypothesis development. The model presented is also interesting due to the fact that an intrinsic moment of momentum or a particle spin naturally occurs in it in 2D and 3D cases with the minimum energy [10].

%********************Bibliography.*******************************************************

\end{document}